\documentstyle[11pt,aaspp4]{article}
\slugcomment{Submitted to Astron. J.}
\lefthead{Haisch et al.}
\righthead{A Near-Infrared Multiplicity Survey of Class I/Flat-Spectrum Systems in Six Nearby Molecular Clouds}

\begin{document}

\title{A Near-Infrared Multiplicity Survey of Class I/Flat-Spectrum Systems in Six Nearby Molecular Clouds}

\author{Karl E. Haisch Jr.\altaffilmark{1,2,3}}
\affil{NASA Ames Research Center, Mail Stop 245-6, Moffett Field, California  94035-1000, khaisch@mail.arc.nasa.gov}

\and

\author{Thomas P. Greene\altaffilmark{2}}
\affil{NASA Ames Research Center, Mail Stop 245-6, Moffett Field, California  94035-1000, tgreene@mail.arc.nasa.gov}

\and

\author{Mary Barsony\altaffilmark{4,5}}
\affil{NASA Ames Research Center, Mail Stop 245-6, Moffett Field, California  94035-1000, mbarsony@stars.sfsu.edu}

\and

\author{Steven W. Stahler}
\affil{Astronomy Department, University of California, Berkeley, 601 Campbell Hall, Berkeley, California 94720-3411, sstahler@astro.berkeley.edu}

\altaffiltext{1}{Current Address: Department of Astronomy, University of Michigan, 830 Dennison Bldg., Ann Arbor, MI 48109-1090}

\altaffiltext{2}{Visiting Astronomer at the Infrared Telescope Facility which is
operated by the University of Hawaii under contract to the National Aeronautics
and Space Administration.}

\altaffiltext{3}{Visiting Astronomer, Cerro Tololo Inter-American Observatory, National Optical Astronomy Observatories, operated by the Association of Universities for Research in Astronomy (AURA), Inc., under cooperative agreement with the National Science Foundation.}

\altaffiltext{4}{Space Science Institute, 4750 Walnut Street, Suite 205, Boulder, CO 80301}

\altaffiltext{5}{NASA Faculty Fellow}

\begin{abstract}

We present new near-IR observations of 76 Class I/flat-spectrum objects in the nearby ($d$ $\la$ 320 pc) Perseus, Taurus, Chamaeleon I and II, $\rho$ Ophiuchi, and Serpens dark clouds. These observations are part of a larger systematic infrared multiplicity survey of self-embedded objects in the nearest dark clouds. When combined with the results of our previously published near-infrared multiplicity survey, we find a $restricted$ companion star fraction of 14/79 (18\% $\pm$ 4\%) of the sources surveyed to be binary or higher order multiple systems over a separation range of $\sim$300 -- 2000 AU with a magnitude difference $\Delta$$K$ $\leq$ 4, and with no correction for background contamination or completeness. This is consistent with the fraction of binary/multiple systems found among older pre-main-sequence T Tauri stars in each of the Taurus, $\rho$ Ophiuchi, and Chamaeleon star-forming regions over a similar separation range, as well as the combined companion star fraction for these regions. However, the companion star fraction for solar-type, and lower mass M dwarf, main-sequence stars in the solar neighborhood in this separation range (11\% $\pm$ 3\%) is approximately one-half that of our sample. Together with multiplicity statistics derived for previously published samples of Class 0 and Class I sources, our study suggests that a significant number of binary/multiple objects may remain to be discovered at smaller separations among our Class I/flat-spectrum sample and/or most of the evolution of binary/multiple systems occurs during the Class 0 phase of early stellar evolution.

\end{abstract}


\section{Introduction}

Traditional multiplicity search techniques such as direct imaging and spectroscopy have revealed that most field stars are members of binary or higher order multiple systems (\cite{al76}; \cite{dm91}; \cite{fm92}). Infrared surveys of younger pre-main-sequence (PMS) T Tauri stars (TTSs) in nearby, young dark cloud complexes (i.e., Taurus, Ophiuchus, Chamaeleon, Lupus, Corona Australis) conducted over the past ten years have shown that the fraction of binary and multiple stars is even higher in these regions (\cite{gnm93}; \cite{math94}; \cite{simon95}; \cite{gmpb97}; \cite{bkm03}). In contrast, the results of multiplicity studies of the young Trapezium and NGC 2024 clusters in Orion are consistent with what one finds for field stars (\cite{ms94}; \cite{pro94}; \cite{petr98}; \cite{scb99}; \cite{bsc03}). 

Each of the above studies samples a different range of separations and companion brightness, making subsequent intercomparison difficult. However, using a model which accounts for differences in sample completeness, dynamic range, and chance projection of background objects, Duch\^{e}ne (1999) has reanalyzed and confirmed the previous findings. Thus, the formation of binary and multiple systems appears to be the rule, rather than the exception, in star-forming regions.

However, very little is currently known about the multiplicity of even younger, self-embedded young stellar objects (YSOs). Recently, Reipurth (2000) analyzed the multiplicity of 14 young (ages $\leq$ 10$^{5}$ yr) sources which drive giant Herbig-Haro flows. Between 79\% (separation = $\sim$ 10 -- 3700 AU)  and 86\% (separation = $\sim$ 10 -- 5000 AU) of these sources had at least one companion and, of these, half were higher order multiple systems. In addition, in a millimeter survey of young embedded objects, Looney, Mundy, \& Welch (2000) find a very high multiplicity fraction. All surveyed objects were members of small groups or binary systems. These fractions are even larger than those found among either the PMS TTS or field star populations. In contrast, a significantly lower frequency (36\% $\pm$ 14\%; separation = $\sim$ 300 -- 2000 AU) of binary/multiple stars was observed in a sample of 19 Class I and flat-spectrum YSOs in the $\rho$ Ophiuchi and Serpens dark clouds (\cite{hbgr02}). One needs to know with reasonable statistical confidence the multiplicity properties of embedded Class I/flat-spectrum sources, which previous surveys do not provide.

In this paper, we present the results of a new near-infrared imaging survey of 76 self-embedded young stars in the $\rho$ Ophiuchus ($d$ = 125 pc; \cite{kh98}), Serpens ($d$ = 310 pc; \cite{del91}), Taurus ($d$ = 140 pc; \cite{kh95}), Perseus ($d$ = 320 pc; \cite{her98}), and Chamaeleon I and II ($d$ = 160 pc and $d$ = 178 pc respectively; \cite{whi97}) star-forming regions. All of the sources in our survey were selected such that they have either Class I or flat spectral energy distributions as determined from previous studies using IRAS, ISO, and ground-based  data (e.g., \cite{wly89}; \cite{pww92}; \cite{llm93}; \cite{greene94}; \cite{kh95}; \cite{per00}; \cite{bon01}; \cite{kaas01}). In defining Class I/flat-spectrum YSOs, the classification scheme of Greene et al. (1994) has been adopted as it is believed to correspond well to the physical stages of evolution of YSOs (e.g. \cite{am94}). Thus, Class I sources have a spectral index [$\alpha$ = dlog($\lambda$F$_{\lambda}$)/dlog($\lambda$)] $>$ 0.3, and flat-spectrum sources have 0.3 $>$ $\alpha$ $\geq$ --0.3 in the wavelength range 2 $\mu$m $\le$ $\lambda$ $\le$ 10 $\mu$m. Combined with the results of our previously published Class I/flat-spectrum multiplicity survey (\cite{hbgr02}), our sample includes 9/21 published known Class I/flat-spectrum sources in Perseus, 11/21 in Taurus, and all known Class I/flat-spectrum YSOs in $\rho$ Ophiuchi, Serpens, and Chamaeleon I and II.

We discuss our near-IR observations and data reduction procedures in $\S$2. In $\S$3, we present the results of our survey, and discuss the results in $\S$4. We summarize our primary results in $\S$5.

\section{Observations and Data Reduction}

\subsection{OSIRIS-NSFCAM Observations}

All near-infrared observations were obtained using two separate facilities. The near-IR $JHKL$-band (1.25, 1.65, 2.2, and 3.5 $\mu$m) observations of all Taurus, Perseus, and Serpens (and follow-up $L$-band data for $\rho$ Ophiuchi) Class I/flat-spectrum sources were obtained with the NSFCAM 256$\times$256 pixel InSb facility array camera on the NASA IRTF 3 m telescope on Mauna Kea, Hawaii (\cite{ray93}; \cite{shu94}). Similar $JHK$ observations of all Chamaeleon I, II and $\rho$ Ophiuchi sources were taken with OSIRIS, the Ohio State 1024$\times$1024 pixel HAWAII HgCdTe infrared imager/spectrometer array on the Cerro Tololo Inter-American Observatory (CTIO) 4 m telescope (\cite{atw92}; \cite{dep93}).

The $J, H$, and $K$-band observations of the Chamaeleon I, II, and $\rho$ Ophiuchi sources were made during the period 2002 Feb 28 - Mar 3. For sources found to be single, only $K$-band data were obtained. The plate scale of OSIRIS on the CTIO 4 m telescope is 0\farcs161 pixel$^{-1}$ with a corresponding field of view of approximately 93\arcsec$\times$93\arcsec. The average FWHM for all observations was $\sim$ 0\farcs6. Each source was observed in a five point dither pattern (a 2 $\times$ 2 square with a point at the center) with 12\arcsec \hspace*{0.05in}offsets between the corners of the square. Typical total integration times ranged from 1 - 5 minutes at $J$ and $H$-bands, and 2 to 3 minutes at $K$-band. These integration times yield 5$\sigma$ sensitivity limits, good to within 0.2 magnitudes, of $\sim$ 20.0 - 20.7 magnitudes at $J$-band, $\sim$ 19.0 - 19.7 magnitudes at $H$-band, and $\sim$ 18.5 at $K$-band.

Similar $J, H, K$, and $L$-band observations of all Taurus, Perseus, and Serpens sources (as well as $L$-band observations of all $\rho$ Ophiuchi sources for which $JHK$ data were obtained at CTIO as described above) were made during the periods 2001 Dec 08 - 10, 2002 Jan 03 - 05, and 2002 Jun 13 - 15 using NSFCAM at the NASA IRTF 3 m telescope. Again, for single sources, only $K$-band data were taken. For this study, we used a plate scale of 0\farcs148 pixel$^{-1}$ with a corresponding field of view of approximately 38\arcsec$\times$38\arcsec. The average FWHM for all observations was $\sim$ 1\arcsec. Each source was observed in a five point dither pattern (a 2 $\times$ 2 square with a point at the center) with 12\arcsec \hspace*{0.05in}offsets between the corners of the square. At each dither position, the telescope was nodded to separate sky positions 30\arcsec \hspace*{0.05in}north of each target observation. Typical total integration times ranged from 5 to 10 minutes at $J$-band, 1 - 2 minutes at $H$ and $K$-bands, and 60 seconds at $L$-band. These integration times yield 5$\sigma$ sensitivity limits, good to within 0.2 magnitudes, of $\sim$ 20.5 - 21.0 magnitudes at $J$-band, $\sim$ 19.0 - 19.3 magnitudes at $H$-band, $\sim$ 18.5 - 19.0 and $K$-band, and $\sim$ 14.0 at $L$-band.

All $JHKL$ data were reduced using the Image Reduction and Analysis Facility (IRAF)\footnote[5]{IRAF is distributed by the National Optical Astronomy Observatories, which are operated by the Association of Universities for Research in Astronomy, Inc., under cooperative agreement with the National Science Foundation.}. The individual sky frames were normalized to produce flat fields for each target frame. All target frames were processed by subtracting the appropriate sky frames and dividing by the flat fields. Finally all target frames were registered and combined to produce the final images of each object in each filter.

\subsection{Source Photometry and Calibration}

Aperture photometry was performed using the PHOT routine within IRAF. An aperture of 4 pixels in radius was used for all target photometry, and a 10 pixel radius was used for the standard star photometry. Sky values around each source were determined from the mode of intensities in an annulus with inner and outer radii of 10 and 20 pixels, respectively. Our choice of aperture size for our target photometry insured that the individual source fluxes were not contaminated by the flux from companion stars, however they are not large enough to include all the flux from a given source. In order to account for this missing flux, aperture corrections were determined using the MKAPFILE routine within IRAF. Aperture photometry was performed on all target sources using the same 10 pixel aperture used for the photometry of the standard stars. Fluxes in both the 10 pixel and 4 pixel apertures were compared, and the instrumental magnitudes for all sources were corrected to account for the missing flux.

Photometric calibration was accomplished using the list of standard stars of Elias et al. (1982) for all IRTF data, and the HST/NICMOS infrared standard stars of Persson et al. (1998) for the CTIO observations. The standards were observed on the same nights and through the same range of airmasses as the target sources. Zero points and extinction coefficients were established for each night. All NSFCAM magnitudes and colors were transformed to the CIT system using Mauna Kea to NSFCAM and NSFCAM to CIT photometric color transformation equations from http://irtf.ifa.hawaii.edu/Facility/nsfcam/mkfilters.html, http://irtf.ifa.hawaii.edu/Facility/nsfcam/hist/color.html and the NSFCAM User's Guide. Following the expectations discussed in the OSIRIS User's Manual, we transformed all OSIRIS magnitudes and colors to the CIT system using the Cerro Tololo Infrared Imager (CIRIM) to CIT transformation equations from the CIRIM Manual.  The photometric uncertainty for all observations is typically good to within $\pm$ 0.10, 0.04, 0.02, and 0.06 magnitudes at $J$, $H$, $K$, and $L$-band respectively.

\section{Analysis and Results}

The companion star fraction (CSF) is defined as:

\begin{equation}
CSF = \frac {\it B + 2T + 3Q}{\it S + B + T + Q}
\end{equation}

\noindent where $S$ is the number of single stars, $B$ is the number of binary systems, $T$, the number of triple systems, and $Q$, the number of quadruple systems. In Table~\ref{table1}, we summarize, for each region surveyed, the minimum projected separation to which our observations are sensitive, the number of sources observed, the number of sources found to be binary/multiple, and the CSFs. The quoted uncertainties in the CSFs represent the statistical standard errors (i.e., $\sqrt{[CSF(1 - CSF)]/N}$). Our lower limit for detectable separations of 100 -- 140 AU for Taurus, Chamaeleon, and $\rho$ Ophiuchi and $\sim$ 300 AU for Perseus and Serpens is set by the seeing ($\sim$ 0\farcs6 at CTIO and $\sim$ 1\arcsec \hspace*{0.05in}at the IRTF). We have imposed an upper limit to the separations of 2000 AU (corresponding to $\la$ 6\farcs3 in Perseus and Serpens, $\la$ 11\arcsec \hspace*{0.05in}in Chamaeleon, $\la$ 14\arcsec \hspace*{0.05in}in Taurus, and $\la$ 16\arcsec \hspace*{0.05in}in $\rho$ Ophiuchi) in order to avoid including sources which are not gravitationally bound systems (e.g., \cite{rz93}; \cite{simon95}). Within the errors, the CSFs are the same for each region, with the exception of Taurus, in which no binary/multiple objects were detected among the surveyed objects. However, it remains possible that the Perseus and Chamaeleon regions may have somewhat lower CSFs ($\sim$ 10\% -- 20\%) than those found in Serpens and $\rho$ Ophiuchi ($\sim$ 25 -- 35\%), although this is only marginally significant given the large error bars on the calculated CSFs.

The CSFs quoted in Table~\ref{table1} assume that there are no restrictions on either separations between companions or magnitude differences between components, an assumption that is clearly unattainable in practice. The quantity that one can measure is a $restricted$ CSF, that is the CSF over a given physical separation range to a stated component $K$-magnitude difference ($\Delta$$K$). In calculating the overall restricted CSF for our sample, we restrict the physical separation range to 300 -- 2000 AU, since 300 AU corresponds to our lower limit for detectable separations in Perseus and Serpens. In addition, sensitivity calculations from our data indicate that we can detect a $K$ = 4 magnitude difference between the primary and companion at a separation of 1\arcsec \hspace*{0.05in}at the 5$\sigma$ confidence level. Thus, we restrict our analysis to component magnitude differences $\Delta$$K$ $\leq$ 4 mag. 

In combination with our previously published results for $\rho$ Ophiuchi and Serpens (in which we find a restricted CSF of 5/13; 38\% $\pm$ 13\% over a separation range of 300 -- 2000 AU), we find a $restricted$ CSF of 14/79 (18\% $\pm$ 4\%) for separations between $\sim$ 300 -- 2000 AU and $\Delta$$K$ = 4 mag, with no correction for background contamination or completeness. In the calculation of our restricted CSF, the sources WL 1 and EC 82/EC 86 are excluded since their separations (103 AU and $\sim$ 2700 AU respectively) are outside of our restricted separation range. Furthermore, the sources IRS 54, GY 51, EC 129, GY 91, WL 22, GY 197, EC 121, EC 40, EC 37, EC 28, and DEOS are not included in the calculation of the overall restricted CSF as their component $\Delta$$K$ values are either not within $\Delta$$K$ $\leq$ 4, or any potential companions would be fainter than our 5$\sigma$ sensitivity limit.

Table~\ref{table2} lists the separations (in both arcsec and AU) and position angles (measured with respect to the brightest source at $K$-band) for the sources in our survey which were found to be binary/multiple. In Tables~\ref{table3} --~\ref{table8}, we present the Right Ascension and Declination coordinates (J2000), $K$-band magnitudes, or in the case of the multiple objects in Table~\ref{table3}, $JHKL$ magnitudes and near-IR colors, for all surveyed sources (the exception being in Chamaeleon, where $L$-band data were not taken).

None of the sources in the Ced110 IRS6, HB 1, and GCNM 53 systems were detected at $J$-band, and neither component in the HB 1 and GCNM 53 systems were detected at $H$-band. This was not likely due to the smearing out of faint sources in these relatively long (90 second) exposures, since this effect was not observed in other equally long $J$ and $H$-band images in which faint sources were detected. Furthermore, the mean object sizes were the same in all bands. Upper limits for the $J$ and $H$-band magnitudes for these sources were determined by adding artificial stars to the respective $J$ and $H$-band images, and counting the number of sources recovered by DAOFIND. Artificial stars were added at random positions to each image in twenty separate half magnitude bins with each bin containing one hundred stars. The twenty bins covered a magnitude range from 15.0 to 25.0. The artificial stars were examined to ensure that they had a similar FWHM of the point-spread function as the sources detected in other $J$ and $H$-band images. Aperture photometry was performed on all sources to confirm that the assigned magnitudes of the added sources agreed with those returned by PHOT. All photometry agreed to within 0.1 magnitudes. DAOFIND and PHOT were then run and the number of identified artificial sources within each half magnitude bin was tallied. This process was repeated 20 times. For Ced 110 IRS6, HB 1, and GCNM 53, our 5$\sigma$ $J$-band magnitude limit, good to within $\pm$ 0.2 magnitudes, is 20.0. Similarly, our 5$\sigma$ $H$-band magnitude limit is 19.0. We list our 5$\sigma$ $J$ and $H$-band limits in Table~\ref{table3} where appropriate.

\section{Discussion}

\subsection{Multiplicity Characteristics}

For the ClassI/flat-spectrum sources surveyed, we find a $restricted$ CSF of 14/79 (18\% $\pm$ 4\%). This consistent with the restricted CSFs derived for PMS T Tauri stars in each of the Taurus, $\rho$ Ophiuchi, and Chamaeleon star-forming regions over a separation range of $\sim$ 300 -- 1800 AU (\cite{lei93}; \cite{simon95}; \cite{gmpb97}; \cite{all02}; \cite{bkm03}), as well as the combined restricted CSF (19\% $\pm$ 3\%) for these regions. In contrast, however, the CSF for solar-type, and lower mass M dwarf, main-sequence stars in the solar neighborhood in this separation range (11\% $\pm$ 3\%; \cite{dm91}; \cite{fm92}) is approximately one-half that of our sample.

Reipurth (2000) found that 79\% -- 86\% of young (ages $\leq$ 10$^{5}$ yr) stellar objects driving giant Herbig-Haro flows were multiple, with half being triple or even higher-order systems. This CSF is considerably higher than the restricted CSF found in the present study. However, the Reipurth work was based on adaptive optics, HST NICMOS, and VLA data, all having higher spatial resolution than our present Class I/flat-spectrum study. We have determined that we would have detected 4 of these sources as binaries (no triples) given the spatial resolution, dynamic range, and physical separation limits of our present survey. This results in a restricted CSF = 29\% $\pm$ 12\%, statistically identical to the restricted CSF of our Class I/flat-spectrum sample. 

In a $\lambda$ = 2.7 mm interferometric survey of 24 YSOs, Looney, Mundy, \& Welch (2000) found that all of the embedded objects were members of small groups or binary systems. All but four of these objects are Class 0 or Class I sources. Together with the multiplicity statistics found among older T Tauri and main sequence stars, our study suggests that a significant number of binary/multiple objects may remain to be discovered at smaller separations among our Class I/flat-spectrum YSOs and/or most of the evolution of binary/multiple systems occurs during the Class 0 phase of early stellar evolution.

\subsection{Notes on the Multiple Sources}

{\em 03260$+$3111} --- This source forms a wide binary system with a 3\farcs62 separation at P.A. $\simeq$ 48$^\circ$. Clark (1991) determined fluxes in all four Infrared Astronomical Satellite (IRAS) bands for 03260$+$3111, although it was noted that several sources were present in the IRAS images. Subsequently, Ladd, Lada, \& Myers (1993) obtained $H$ and $K$-band photometry of this source and determined a total luminosity of L = 318 L$_{\odot}$. Based on an optical and near-infrared imaging survey, Magnier et al. (1999) have classified 03260$+$3111 as a transitional YSO (i.e., flat-spectrum), consistent with the near-infrared colors derived in the present survey.

{\em ChaI T33} --- Also known as Glass 1, ChaI T33 (catalog designation from Whittet et al. 1987) was first found to exhibit a strong near-infrared excess by Glass (1979). Chelli et al. (1988) identified ChaI T33 as a binary source (separation = 2\farcs67; P.A. 285$^\circ$) for the first time using near-infrared narrow slit scan observations. Most of the total system luminosity (L $\simeq$ 5 L$_\odot$) was found to be associated with the very red companion, while the optically dominant primary is a non-emission line K4 star. Since the companion is much brighter in the near-infrared than the primary, the IRAS fluxes of T33 have been assigned to this component (\cite{pww92}). The binarity of ChaI T33 has also been noted by Feigelson \& Kriss (1989) and Reipurth \& Zinnecker (1993). The former authors have found ChaI T33 to be an Einstein X-ray source (CHX 12), with the red companion being a weak-line H$\alpha$ spectral type G5 star. Model fits to the spectral energy distributions (SEDs) of both components by Koresko, Herbst, \& Leinert (1997) suggest that the primary is consistent with its classification as a ``naked'' T Tauri star by Feigelson \& Kriss (1989) with an implied mass of M = 1 M$_\odot$ and an age of 3 $\times$ 10$^{6}$ yr. The observed spectral type of the companion was also found to be consistent with that derived from the SED modeling. Finally, ISOPHOT spectra suggest a strong mid-infrared variability of ChaI T33 (\cite{gur99}), already suggested by a variability flag in the IRAS fluxes of this object. Comparably strong near-infrared variability was also noted previously by Chelli et al. (1988), who concluded that the primary is the variable component. The near-infrared variability of T33 has also been noted more recently by Carpenter et al. (2002). The ISOPHOT spectra of G\"{u}rtler et al. (1999) suggest that the primary may also be variable at mid-infrared wavelengths or that the infrared companion is a variable also.

{\em Ced110 IRS6} --- Detected as an embedded IRAS source by Prusti et al. (1991), Ced110 IRS6 is located in the direction of a dense molecular core discovered by Mattila, Liljestr\"{o}m, \& Toriseva (1989) at the edge of the Cederblad 110 optical reflection nebula. Using IRAS fluxes in conjunction with near-infrared photometry, Prusti et al. (1991) have classified IRS6 as a Class I YSO with a luminosity of L = 1 L$_\odot$. Ced110 IRS6 was detected as an X-ray source with L$_{X}$ $\simeq$ 10$^{29 - 30}$ ergs s$^{-1}$ in archived ROSAT PSPC pointed observations (\cite{ckf98}), and is thus one of the lowest X-ray luminosity Class I objects yet detected. Ced110 IRS6 has also been detected as an unresolved mid-infrared excess source in ISOCAM images of the Ced110 dense core (\cite{per00}). More recently, sub-arcsecond near-infrared images of IRS6 have revealed the binarity of this object (\cite{per01}). In contrast to the previous system luminosity determination of Prusti et al. (1991), the primary component is found to have a luminosity of L = 6 L$_\odot$, with the secondary source having a luminosity about a factor of 7 lower, although with similar $H$ -- $K$ colors. This suggests that most of the observed mid- and far-infrared ISOCAM and ISOPHOT (\cite{leh01}) flux densities are associated with the primary component of Ced110 IRS6.

{\em ChaI T14a} --- ChaI T14a is also known as Herbig-Haro (HH) 48. The binary nature of this object was suggested initially by Schwartz (1977), who noted that HH 48 consisted of two close condensations, the northeast component being the fainter. Indeed, our resolved (separation = 2\farcs35; P.A. = 52$^\circ$) near-infrared photometry of ChaI T14a confirms that the secondary (northeast) source is the fainter component of the binary. A subsequent spectrum of ChaI T14a implied that it is a T Tauri star with relatively strong forbidden emission lines (\cite{sjs84}). ChaI T14a has been been detected at both mid- and far-infrared wavelengths by ISOCAM and IRAS respectively (\cite{pww92}; \cite{per01}), and exhibits a flat-spectrum composite SED.

{\em IRS 48} --- IRS 48 forms a wide ($\sim$ 15\arcsec) binary with IRS 50. IRS 48 has been classified as a Class I YSO (\cite{wly89}; hereafter WLY), although recent ISOCAM observations suggest a flat-spectrum object (\cite{bon01}). IRS 50 remains unclassified, however our $JHKL$ imaging data suggest infrared colors consistent with a Class II YSO. IRS 48 and IRS 50 exhibit very different spectral types of $<$F3 and M4 respectively (\cite{lr99}). IRS 48 has been detected at 1.3 mm (\cite{am94}), and is also a CO outflow driving source exhibiting a bipolar morphology and high-velocity wings (\cite{bon96}).

{\em IRS 54} --- We find IRS 54 to be a binary source with a separation of $\simeq$ 7\farcs3 at a position angle of 324$^\circ$. IRS 54 itself is a Class I YSO (WLY; \cite{bon01}), while the companion has $JHKL$ colors indicative of a heavily reddened (A$_{v}$ $>$ 25) Class III or background object. IRS 54 was detected as a ROSAT x-ray source with L$_{x}$ = 10$^{30.7}$ erg s$^{-1}$ (\cite{cas95}), however subsequent ROSAT PSPC (\cite{ckf98}) and Chandra x-ray (\cite{ikt01}) observations resulted in non-detections.

{\em GY 51} --- GY 51 was first discovered to be a binary object with a separation of 1\farcs15 (very similar to our derived separation of 1\farcs16) by Strom, Kepner, \& Strom (1995). We find GY 51 to be in fact a triple system with a separation of 5\farcs66 and a position angle of 97.7 degrees. ISOCAM observations reveal GY 51 to be a flat-spectrum source (\cite{bon01}), however its dereddened spectral index suggests a Class II YSO (\cite{wilk01}). GY 51 is also a known radio source (LFAM 9; \cite{lfam91}).

{\em WL 2} --- WL 2 was discovered by Wilking \& Lada (1983) and was subsequently found to be binary by Barsony et al. (1989). Barsony, Ressler, \& Marsh (2003) find a spectral index of $+$0.21 for the primary, and -0.46 for the secondary, making this a flat-spectrum and Class II pairing. WL 2 has been detected at 1.3 mm (\cite{am94}) and is a Chandra x-ray source (\cite{ikt01}).

{\em IRS 37} --- IRS 37 (also known as GY 244) forms a wide ($\sim$ 8\farcs5) binary with WL 5 (= IRS 38), and is part of an asterism which also includes WL 3 and WL 4. Both objects were discovered in IRAS and near-infrared observations of WLY and Wilking \& Lada (1983) respectively. Barsony et al. (1989) subsequently found the near-infrared counterpart to IRS 37. The exact spectral classification of IRS 37 has been the subject of some debate. WLY cite IRS 37 to be a Class I YSO, a classification supported by Bontemps et al. (2001), while Andr\'{e} \& Montmerle (1994) assign a Class II designation. WL 5, however, is considered to be an extinguished Class III YSO (\cite{am94}, and references therein; \cite{bon01}). Indeed, it is one of the most heavily extincted infrared sources in the $\rho$ Oph cloud core (\cite{adpl92}). As was the case with IRS 48/IRS 50, IRS 37 and WL 5 have very different spectral types of M4 (\cite{lr99}) and F7 (\cite{gm95}) respectively . Both objects are known x-ray flare sources, however the flares observed in IRS 37 are significantly stronger (\cite{ikt01}).

{\em EC 129} --- EC 129 appears to be the brightest component of a wide binary system, although the physical separation (1930 AU) is almost at our imposed upper limit for a bound system. First detected by Eiroa \& Casali (1992), EC 129 exhibits 3.08$\mu$m H$_{2}$O-ice absorption ($\tau$ = 0.08) and has an extinction corrected luminosity of L = 3.29 L$_\odot$. Casali \& Eiroa (1996) found that its 2 $\mu$m $\delta v = 2$ CO bands were in absorption with a CO index of 2.4. EC 129 may be associated with the IRAS source PS 2 identified in Hurt \& Barsony (1996). This source exhibits polarization consistent with scattering in the bipolar/cometary nebula associated with the object (\cite{sog97}). Testi \& Sargent (1998) detected 3 mm continuum emission from EC 129, and ISO-LWS spectra have identified EC 129 as a point source of $\sim$ 5\arcsec size at a dust temperature of T$_{d} \simeq$ 24 K (\cite{lar00}).

{\em HB 1} --- HB 1 was identified as an IRAS source by Hurt \& Barsony (1996), and is coincident with a 3 mm continuum source detected by Testi \& Sargent (1998). Kaas (1999) identified a near-infrared source associated with HB 1, which we have subsequently found to be binary (separation = 1\farcs46; P.A. = 29$^\circ$) in the present study.

{\em GCNM 53} --- Identified in a deep optical and infrared imaging survey by Giovannetti et al. (1998), we find GCNM 53 to be a wide binary with a separation of 4\farcs31 at a position angle of 159$^\circ$.

\section{Summary and Conclusions}

\noindent\large$\bullet$\normalsize\hspace*{0.1in}We have obtained new near-IR observations of 76 Class I/flat-spectrum objects in the Perseus, Taurus, Chamaeleon I and II, $\rho$ Ophiuchi, and Serpens dark clouds. The observations presented here are part of a larger systematic infrared multiplicity survey of self-embedded objects in the nearest dark clouds.

\noindent\large$\bullet$\normalsize\hspace*{0.1in}When combined with our previously published Class I/flat-spectrum multiplicity survey, we find a $restricted$ CSF of 14/79 (18\% $\pm$ 4\%) over a separation range of $\sim$300 -- 2000 AU and $\Delta$$K$ $\leq$ 4 magnitudes. This is consistent with the restricted CSFs derived for T Tauri stars in each of the Taurus, $\rho$ Ophiuchi, and Chamaeleon star-forming regions over a similar separation range, as well as the combined restricted CSF (19\% $\pm$ 3\%) for these regions. In contrast, the CSF for solar-type, and lower mass M dwarf, main-sequence stars in the solar neighborhood in this separation range (11\% $\pm$ 3\%) is approximately one-half that of our sample.

\noindent\large$\bullet$\normalsize\hspace*{0.1in}Accounting for differences in spatial resolution between the present study and that of Reipurth (2000), we find the restricted CSFs for our Class I/flat-spectrum sample and young (ages $\leq$ 10$^{5}$ yr) stellar objects driving Herbig-Haro flows to be statistically identical. Furthermore, in a millimeter survey of young embedded objects, Looney, Mundy, \& Welch (2000) found that all of the Class 0 and Class I objects surveyed were members of small groups or binary systems. When compared with previously published results, our survey suggests that a significant number of binary/multiple objects may remain to be discovered at smaller separations among our Class I/flat-spectrum YSOs and/or most of the evolution of binary/multiple systems occurs during the Class 0 phase of early stellar evolution.

\acknowledgements

We thank the referee for providing helpful suggestions that improved the manuscript. We thank the CTIO and IRTF staff for their outstanding support in making our observations possible. We also wish to thank Amanda Kaas for providing near-IR data for Class I/flat-spectrum sources in Serpens in advance of publication. K. E. H. gratefully acknowledges support from a National Research Council Research Associateship Award. T. P. G. acknowledges grant support from the NASA Origins of Solar Systems Program, NASA RTOP 344-37-22-11. M. B. gratefully acknowledges NSF grant AST-0206146 which made her contributions to this work possible. Additional support for this work was provided by the National Aeronautics and Space Administration through Chandra Award Number AR1-2005A and AR1-2005B issued by the Chandra X-Ray Observatory Center, which is operated by the Smithsonian Astrophysical Observatory for and on behalf of NASA under Contract NAS8-39073. S. W. S. acknowledges support through NSF grant AST 99-87266.

\newpage
\clearpage
\begin{deluxetable}{lcccr}
\footnotesize
\tablecaption{Companion Star Fraction for Each Region \label{table1}}
\tablewidth{0pt}
\tablehead{Region & Separation (AU)\tablenotemark{a} & \# Observed\tablenotemark{b} & \#
Binary/Multiple\tablenotemark{c} & CSF (\%)}
\startdata
Perseus & 320 & 9 & 1 & 11\% $\pm$ 10\%\nl 
 & \nl
Taurus & 140 & 11 & 0 & 0\% $\pm$ 0\%\nl 
 & \nl
Chamaeleon I \& II & 100 & 17 & 3 & 18\% $\pm$ 9\%\nl 
 & \nl
Serpens & 310 & 19 & 6 & 32\% $\pm$ 11\%\nl 
 & \nl
Rho Oph & 125 & 37 & 9 & 27\% $\pm$ 7\%\nl
\enddata
\tablenotetext{a}{Denotes the minimum separation to which our observations are
sensitive.}
\tablenotetext{b}{Number of sources observed in each region.}
\tablenotetext{c}{Number of sources found to be binary/multiple.}
\end{deluxetable}

\clearpage
\begin{deluxetable}{lcccc}
\footnotesize
\tablecaption{Separations and Position Angles for the Binary/Multiple Sources \label{table2}}
\tablewidth{0pt}
\tablehead{Source & Region & Separation (\arcsec) & Separation (AU) & Position Angle
(degrees)\tablenotemark{a}}
\startdata
03260$+$3111 & Perseus & 3.62 & 1160 & 47.9\nl 
 & & & & \nl
ChaI T33B & Chamaeleon I & 2.38 & 380 & 285.2\nl 
 & & & & \nl
Ced110 IRS6 & Chamaeleon I & 1.95 & 310 & 95.6\nl 
 & & & & \nl
ChaI T14a & Chamaeleon I & 2.35 & 375 & 51.9\nl
 & & & & \nl
IRS 48 & $\rho$ Oph & 15.13 & 1890 & 121.1\nl
 & & & & \nl
IRS 54 & $\rho$ Oph & 7.25 & 905 & 324.0\nl
 & & & & \nl
GY 51 & $\rho$ Oph & 1.16 & 145 & 67.5\nl 
Third source & & 5.66 & 710 & 97.7\nl
 & & & & \nl
WL 2 & $\rho$ Oph & 4.15 & 520 & 343.2\nl
 & & & & \nl
IRS 37 & $\rho$ Oph & 8.55\tablenotemark{b} & 1070 & 65.8\tablenotemark{b}\nl
 & & & & \nl
IRS 43/GY 263 & $\rho$ Oph & 6.99 & 875 & 322.0\nl 
 & & & & \nl
WL 1 & $\rho$ Oph & 0.82 & 103 & 321.2\nl 
 & & & & \nl
GY 23/GY 21 & $\rho$ Oph & 10.47 & 1310 & 322.6\nl 
 & & & & \nl
L1689 SNO2 & $\rho$ Oph & 2.92 & 365 & 240.3\nl 
 & & & & \nl
SVS 20 & Serpens & 1.51 & 468 & 9.9\nl
 & & & & \nl
EC 95/EC 92 & Serpens & 5.03 & 1560 & 352.1\nl 
 & & & & \nl
EC 129 & Serpens & 6.22 & 1930 & 323.4\nl
 & & & & \nl
HB 1 & Serpens & 1.46 & 450 & 28.9\nl
 & & & & \nl
GCNM 53 & Serpens & 4.31 & 1340 & 159.4\nl
\enddata
\tablenotetext{a}{Measured with respect to the brightest source at $K$-band}
\tablenotetext{b}{IRS 37 forms a binary with WL 5. Separation and Position Angle are measured with respect
to IRS 37, the Class I YSO.}
\end{deluxetable}

\clearpage
\begin{deluxetable}{lccrrrrrrr}
\footnotesize
\tablecaption{Positions and $JHKL$ Magnitudes and Colors for Multiple Sources
\label{table3}}
\tablewidth{0pt}
\tablehead{Source & RA(J2000) & Dec(J2000) & J & H & K & L\tablenotemark{a} & (J-H) & (H-K) &
(K-L)\tablenotemark{a}}
\startdata
03260$+$3111 & 03 29 10.40 & $+$31 21 58.0 & 9.36 & 8.08 & 7.29 & 6.75 & 1.28 & 0.79 & 0.54\nl
 & & & 13.61 & 12.07 & 11.04 & 10.22 & 1.54 & 1.03 & 0.82\nl
 & & & & & & & & & \nl
ChaI T33B & 11 08 15.69 & -77 33 47.1 & 9.32 & 8.10 & 6.93 & & 1.22 & 1.17 & \nl
 & & & 9.98 & 9.16 & 8.85 & & 0.82 & 0.31 & \nl
 & & & & & & & & & \nl 
Ced110 IRS6 & 11 07 09.80 & -77 23 04.4 & $>$20.00 & 15.18 & 10.86 & & $>$4.82 & 4.32 & \nl
 & & & $>$20.00 & 17.57 & 12.86 & & $>$2.43 & 4.71 & \nl
 & & & & & & & & & \nl
ChaI T14a & 11 04 24.32 & -77 18 07.2 & 16.42 & 14.32 & 12.54 & & 2.10 & 1.78 & \nl
 & & & 18.43 & 15.66 & 13.85 & & 2.77 & 1.81& \nl
 & & & & & & & & & \nl
IRS 48 & 16 27 37.20 & -24 30 34.0 & 10.40 & 8.72 & 7.71 & 6.12 & 1.68 & 1.01 & 1.59\nl
IRS 50 & 16 27 38.10 & -24 30 40.0 & 12.54 & 10.94 & 9.92 & 9.24 & 1.60 & 1.02 & 0.68\nl
 & & & & & & & & & \nl   
IRS 54 & 16 27 51.70 & -24 31 46.0 & 16.38 & 12.22 & 10.15 & 7.74 & 4.16 & 2.07 & 2.41\nl
 & & & $>$20.00 & 16.13 & 14.29 & 13.14 & $>$3.87 & 1.84 & 1.15\nl
 & & & & & & & & & \nl 
GY 51 & 16 26 30.49 & -24 22 59.0 & 16.41 & 12.22 & 10.16 & 8.16 & 4.19 & 2.06 & 2.00\nl
 & & & 16.98 & 12.99 & 11.09 & 9.42 & 3.99 & 1.90 & 1.67\nl
 & & & $>$20.00 & 17.27 & 15.51 & $>$13.50 & $>$2.73 & 1.76 & $<$2.01\nl
 & & & & & & & & & \nl 
WL 2 & 16 26 48.56 & -24 28 40.4 & 19.26 & 13.95 & 11.15 & 9.24 & 5.31 & 2.80 & 2.80\nl
 & & & 20.07 & 15.15 & 12.42 & 10.60 & 4.92 & 2.73 & 1.82\nl
 & & & & & & & & & \nl
IRS 37 & 16 27 17.54 & -24 28 56.5 & 19.22 & 14.46 & 10.94 & 8.54 & 4.76 & 3.52 & 2.40\nl
WL 5 & 16 27 18.00 & -24 28 55.0 & $>$20.50 & 15.03 & 10.21 & 7.94 & $>$5.47 & 4.82 & 2.27\nl
 & & & & & & & & & \nl
EC129\tablenotemark{b} & 18 30 02.80 & $+$01 12 28.0 & & & 10.07 & & & & \nl
 & & & & & 15.88 & & & & \nl
 & & & & & & & & & \nl
HB 1 & 18 29 59.50 & $+$01 11 59.0 & $>$20.00 & $>$19.00 & 15.67 & 10.89 & $>$1.00 & $>$3.53 & 4.78\nl
 & & & $>$20.00 & $>$19.00 & 16.98 & 11.77 & $>$1.00 & $>$2.22 & 5.21\nl
 & & & & & & & & & \nl
GCNM 53 & 18 29 52.90 & $+$01 14 56.0 & $>$20.00 & $>$19.00 & 16.75 & 12.95 & $>$1.00 & $>$2.45 & 3.80\nl
 & & & $>$20.00 & $>$19.00 & 18.22 & $>$13.50 & $>$1.00 & $>$0.98 & $<$4.72\nl
\enddata
\tablenotetext{a}{L-band magnitudes, and hence (K -- L) colors, not available for Chamaeleon sources.}
\tablenotetext{b}{EC 129 is a binary source, however no $J, H$, or $L$ observations were taken since the binarity
of this source could not be determined prior to the reduction of the data.}
\end{deluxetable}

\clearpage
\begin{deluxetable}{lccr}
\footnotesize
\tablecaption{Positions and $K$ Magnitudes for Single Perseus Sources
\label{table4}}
\tablewidth{0pt}
\tablehead{Source & RA(J2000) & Dec(J2000) & $K$ Mag.}
\startdata
03382$+$3145 & 03 41 22.70 & $+$31 54 46.0 & 8.22\nl
03259$+$3105 & 03 29 03.70 & $+$31 15 52.0 & 8.53\nl 
03262$+$3114 & 03 29 20.40 & $+$31 24 47.0 & 8.56\nl
03380$+$3135 & 03 41 09.10 & $+$31 44 38.0 & 8.63\nl 
03220$+$3035 & 03 25 09.20 & $+$30 46 21.0 & 10.21\nl
03254$+$3050 & 03 28 35.10 & $+$31 00 51.0 & 10.28\nl
03445$+$3242 & 03 47 41.60 & $+$32 51 43.5 & 11.41\nl 
03439$+$3233 & 03 47 05.00 & $+$32 43 09.0 & 12.73\nl
\enddata
\end{deluxetable}

\clearpage
\begin{deluxetable}{lccr}
\footnotesize
\tablecaption{Positions and $K$ Magnitudes for Single Taurus Sources
\label{table5}}
\tablewidth{0pt}
\tablehead{Source & RA(J2000) & Dec(J2000) & $K$ Mag.}
\startdata
Haro 6-13 & 04 32 15.61 & $+$24 29 02.3 & 7.77\nl
GV Tau B & 04 29 23.61 & $+$24 34 06.8 & 7.86\nl 
Haro 6-28 & 04 35 55.87 & $+$22 54 35.5 & 9.60\nl
04489$+$3042 & 04 52 06.90 & $+$30 47 17.0 & 9.98\nl 
04016$+$2610 & 04 04 42.85 & $+$26 18 56.3 & 10.23\nl
04108$+$2803 & 04 13 52.90 & $+$28 11 23.0 & 10.25\nl
04361$+$2547 & 04 39 13.87 & $+$25 53 20.6 & 10.32\nl 
04365$+$2535 & 04 39 35.01 & $+$25 41 45.5 & 10.80\nl
04295$+$2251 & 04 32 32.10 & $+$22 57 30.0 & 11.01\nl
04264$+$2433 & 04 29 30.30 & $+$24 39 54.0 & 11.60\nl
\enddata
\end{deluxetable}

\clearpage
\begin{deluxetable}{lccr}
\footnotesize
\tablecaption{Positions and $K$ Magnitudes for Single Chamaeleon Sources
\label{table6}}
\tablewidth{0pt}
\tablehead{Source & RA(J2000) & Dec(J2000) & $K$ Mag.}
\startdata
ChaI T32 & 11 08 04.61 & -77 39 16.9 & 6.13\nl
ChaI T44 & 11 10 01.35 & -76 34 55.8 & 6.43\nl
ChaI T41 & 11 09 50.39 & -76 36 47.6 & 6.99\nl 
ChaI T42 & 11 09 54 66 & -76 34 23.7 & 7.02\nl
ChaI T29 & 11 07 59.25 & -77 38 43.9 & 7.19\nl
ISO-ChaI 26 & 11 08 04.00 & -77 38 42.0 & 8.25\nl
ChaI C1-6 & 11 09 23.30 & -76 34 36.2 & 8.43\nl
ChaI C9-2 & 11 08 37.37 & -77 43 53.5 & 8.62\nl
ChaII 8 & 12 53 42.88 & -77 15 05.7 & 8.76\nl
ChaI T47 & 11 10 50.78 & -77 17 50.6 & 8.77\nl
ISO-ChaI 97 & 11 07 18.30 & -77 23 13.0 & 11.20\nl
ISO-ChaI 225 & 11 09 55.00 & -76 31 12.0 & 12.38\nl
ISO-ChaI 138 & 11 08 19.20 & -77 30 41.0 & 12.90\nl
ISO-ChaI 86 & 11 06 57.20 & -77 22 51.0 & 13.12\nl 
\enddata
\end{deluxetable}

\clearpage
\begin{deluxetable}{lccr}
\footnotesize
\tablecaption{Positions and $K$ Magnitudes for Single $\rho$ Oph Sources
\label{table7}}
\tablewidth{0pt}
\tablehead{Source & RA(J2000) & Dec(J2000) & $K$ Mag.}
\startdata
Elias 29 & 16 27 09.43 & -24 37 18.5 & 7.54\nl
IRS 42 & 16 27 21.45 & -24 41 42.8 & 8.56\nl 
GSS 30/IRS 1 & 16 26 21.50 & -24 23 07.0 & 9.03\nl
VSSG 18 & 16 27 28.44 & -24 27 21.9 & 9.20\nl 
GSS 26 & 16 26 10.28 & -24 20 56.6 & 9.38\nl 
IRS 34 & 16 27 15.48 & -24 26 40.6 & 10.26\nl 
WL 17 & 16 27 06.79 & -24 38 14.6 & 10.31\nl
WL 12 & 16 26 44.30 & -24 34 47.5 & 10.43\nl
IRS 46 & 16 27 29.70 & -24 39 16.0 & 10.57\nl
WL 6 & 16 27 21.83 & -24 29 53.2 & 10.77\nl
WL 3 & 16 27 19.30 & -24 28 45.0 & 11.50\nl
CRBR 15 & 16 26 19.30 & -24 24 16.0 & 11.73\nl
GY 312 & 16 27 38.91 & -24 40 20.1 & 11.94\nl
CRBR 12 & 16 26 17.30 & -24 23 49.0 & 12.08\nl
GY 344 & 16 27 45.81 & -24 44 53.7 & 12.33\nl
IRS 33 & 16 27 14.60 & -24 26 55.0 & 12.34\nl
GY 245 & 16 27 18.50 & -24 39 15.0 & 12.54\nl
CRBR 85 & 16 27 24.68 & -24 41 03.7 & 14.45\nl
GY 91 & 16 26 40.60 & -24 27 16.0 & 15.76\nl
WL 22 & 16 26 59.30 & -24 35 01.0 & 17.58\nl
GY 197 & 16 27 05.40 & -24 36 31.0 & 18.33\nl
\enddata
\end{deluxetable}

\clearpage
\begin{deluxetable}{lccr}
\footnotesize
\tablecaption{Positions and $K$ Magnitudes for Single Serpens Sources
\label{table8}}
\tablewidth{0pt}
\tablehead{Source & RA(J2000) & Dec(J2000) & $K$ Mag.}
\startdata
EC 94 & 18 29 57.80 & $+$01 12 37.0 & 11.67\nl
EC 73 & 18 29 55.13 & $+$01 13 19.2 & 12.20\nl
EC 38 & 18 29 49.50 & $+$01 17 07.0 & 12.42\nl 
EC 125 & 18 30 02.10 & $+$01 14 00.0 & 12.90\nl 
EC 91 & 18 29 57.80 & $+$01 12 28.0 & 13.67\nl 
EC 80 & 18 29 56.60 & $+$01 12 40.0 & 14.92\nl 
EC 121 & 18 30 01.10 & $+$01 13 26.0 & 15.44\nl
EC 40 & 18 29 49.70 & $+$01 14 57.0 & 15.98\nl 
EC 37 & 18 29 49.10 & $+$01 16 32.0 & 16.01\nl
EC 28 & 18 29 47.00 & $+$01 16 26.0 & 16.47\nl
DEOS & 18 29 49.30 & $+$01 16 19.0 & 16.82\nl
\enddata
\end{deluxetable}

\clearpage

\end{document}